\documentclass[letter]{aa}
\usepackage{tabularx}
\usepackage{amssymb}
\usepackage{amsmath}
\usepackage{txfonts}
\usepackage{balance}
\usepackage{natbib}
\bibpunct{(}{)}{;}{a}{}{,} 
\usepackage{color}
\usepackage{xspace}
\usepackage[normalem]{ulem} 
\usepackage[table]{xcolor} 
\usepackage{multirow}
\usepackage{rotating}
\usepackage{bbm}
\usepackage{cancel}
\usepackage{graphicx}
\usepackage{caption}
\usepackage{ifthen}
\usepackage{booktabs}
\usepackage{tabularx}
\usepackage{float}

\newcommand{\e}[1]{\ensuremath{\times 10^{#1}}}

\makeatletter
\if@referee
    \newcommand{\com}[1]{\textcolor{red}{[#1]}}                            
    \usepackage[nolists,noheads,nomarkers]{endfloat}                       
\else
    \newcommand{\com}[1]{}                                                 
\fi
\makeatother
\usepackage[%
pdfauthor={Fredrik Windmarkl},
pdftitle={Velocity distribution},
pdfstartview=FitH,
linkcolor=blue,
anchorcolor=blue,
citecolor=blue,
filecolor=blue,
menucolor=blue,
urlcolor=blue,
colorlinks=true]{hyperref}

\begin{document}
\title{Breaking through: The effects of a velocity distribution on barriers to dust growth}
\titlerunning{Effects of velocity distributions on barriers to dust growth}
\authorrunning{F. Windmark et al.}
\author{F.~Windmark\inst{1,2} \and T.~Birnstiel\inst{3,4} \and C. W.~Ormel\inst{5,6} \and C. P. Dullemond\inst{1}}
\institute{Universit\"at Heidelberg, Zentrum f\"ur Astronomie, Institut f\"ur Theoretische Astrophysik, Albert-Ueberle-Str. 2, D-69120 Heidelberg, Germany \\e-mail: fwindmark@zah.uni-heidelberg.de \and Member of IMPRS for Astronomy \& Cosmic Physics at the University of Heidelberg \and Excellence Cluster Universe, Boltzmannstr. 2, D-85748 Garching, Germany  \and University Observatory, Ludwig-Maximilians University Munich, Scheinerstr. 1, D-81679, Munich, Germany \and Astronomy Department, University of California, Berkeley, CA 94720, USA \and Hubble Fellow}
\date{Received 13 July 2012 / Accepted 30 July 2012}

\abstract
{It is unknown how far dust growth can proceed by coagulation. Obstacles to collisional growth are the fragmentation and bouncing barriers. However, in all previous simulations of the dust-size evolution, only the mean collision velocity has been considered, neglecting that a small but possibly important fraction of the collisions will occur at both much lower and higher velocities.}
{We study the effect of the probability distribution of impact velocities on the collisional dust growth barriers.}
{We assume a Maxwellian velocity distribution for colliding particles to determine the fraction of sticking, bouncing, and fragmentation, and implement this in a dust-size evolution code. We also calculate the probability of growing through the barriers and the growth timescale in these regimes.}
{We find that the collisional growth barriers are not as sharp as previously thought. With the existence of low-velocity collisions, a small fraction of the particles manage to grow to masses orders of magnitude above the main population.}
{A particle velocity distribution softens the fragmentation barrier and removes the bouncing barrier. It broadens the size distribution in a natural way, allowing the largest particles to become the first seeds that initiate sweep-up growth towards planetesimal sizes.}


\maketitle

\section{Introduction}
\label{sec:introduction}

Primary accretion is the earliest stage of planet formation, where tiny, micrometer-sized dust grains in the protoplanetary disk grow into planetesimals of several kilometers in size. In the classic scenario, this happens by the way of incremental growth, in which sticking collisions lead to successively larger aggregates \citep{1997Icar..127..290W, 2005A&A...434..971D}. As laboratory experiments \citep{2008ARA&A..46...21B, 2010A&A...513A..56G} and numerical simulations \citep{2010A&A...513A..57Z,2010A&A...513A..79B} have improved our understanding of the collision process and the dust evolution, it has become evident that incremental growth to form planetesimals cannot continue unhindered.

As particles grow, they decouple more and more from the surrounding gas, which increases their relative velocities. At the fragmentation barrier, collision energies are high enough to cause particle destruction, halting the dust growth at centimeter to meter sizes \citep{2008A&A...480..859B}. \cite{2010A&A...513A..57Z} also introduced the bouncing barrier, which stops the growth at even smaller sizes. In this case, the collision energies are too low to cause any particle destruction, but also too high for sticking, with growth-neutral bouncing events as the result.

\cite{2012A&A...540A..73W} suggested a sweep-up scenario where the fragmentation barrier can be circumvented. They found that even though collisions between equal-sized particles generally lead to fragmentation, if the mass-ratio is large enough, growth of the larger particle can occur even at very high velocities. In this scenario, the growth initially stalls at the bouncing barrier, but if a small number of slightly larger 'seed' particles are introduced, they rapidly sweep up the smaller particles and grow to very large sizes. The growth barriers in this case limit the number of large particles and therefore reduce the number of destructive collisions among them. Exactly how the first seeds are formed is however still not clear.

All prior dust coagulation models have until now relied on the mean value to describe the velocity at which a collision occurs. In reality, the relative velocities between the particles that arises because of Brownian motion and turbulence does not take a single value, but is better represented by a probability distribution, owing to geometrical and stochastic effects. Here, we focus on turbulence since it is the dominating
source of relative velocity between the small grains below the fragmentation barrier.

A general formula for the probability distribution function (PDF) of particle relative velocities is however unavailable, despite the efforts of many numerical and experimental works. There is tentative evidence that the PDF for particles with large Stokes numbers (${\rm St} \sim 1$, those that couple to the driving scales of the turbulence) is Maxwellian or close to it (\citealt{2010MNRAS.405.2339C}; Dittrich et al., in prep.). However, at smaller sizes (where particles couple to the Kolmogorov scale) the PDF may be better characterized by wide, exponential tails (\citealt{2000JFM...415..117W, 2010JFM...661...73P, Hubbard:2012ud}) Future numerical and analytical modelling is desired to refine and interpret these data. In this work, as a first step, we assume that turbulent velocities are Maxwellian distributed.

A velocity distribution allows some collisions to result in sticking where the average outcome would produce a bouncing or fragmentation event. This causes the barriers to blur out, and might allow for some lucky particles to just by sheer chance repeatedly experience only low-velocity collisions and grow to larger sizes than the main population.

In this letter, we show the effect of such a velocity distribution in a local dust-size evolution code, not only as a method for creating lucky larger particles, but also to see how it affects the general dust population.

\section{Method}
\label{sec:method}

We implement a Maxwellian velocity distribution together with three simple collisions models into a local version of the dust-size evolution code of \cite{2010A&A...513A..79B}. We consider relative velocities arising only from Brownian motion and turbulence, which are the dominant sources for small particles, and for the disk properties, we take the minimum mass solar nebula \citep{1977Ap&SS..51..153W}. It is now well-known known that the gas and dust mass of the solar nebula might have been much higher \citep{2007ApJ...671..878D}, which would increase the coagulation rates and the particle size at the fragmentation barrier \citep{2011A&A...525A..11B}. We refer to Table~\ref{tab:params} for all the parameters used in this work.
\begin{table}
\caption{Disk model parameters used in the local simulations.}
\begin{tabular}{ l l l l }
\hline
\hline
Parameter & Symbol & Value & Unit \\
\hline
Distance to star 	& $r$ 			& 1 				& AU \\
Gas surface density 	& $\Sigma_{\rm g}$ 	& 1700 			& g cm$^{-2}$ \\
Dust surface density & $\Sigma_{\rm d}$ 	& 17 				& g cm$^{-2}$ \\
Gas temperature 	& $T$ 			& 280 			& K \\
Sound speed		& $c_{\rm s}$		& $1.0\cdot10^5$ 	& cm s$^{-1}$ \\
Turbulence parameter & $\alpha$		& $10^{-4}$		& - \\
Solid density of dust grains & $\xi$ 		& 1	 			& g cm$^{-3}$ \\
\hline
\end{tabular}
\label{tab:params}
\end{table}

\subsection{Collision models}

In the simple models presented here, the possible outcome of a collision is one of the three collision types of sticking, fragmentation, and bouncing, so that $p_{\rm s} + p_{\rm f} + p_{\rm b} = 1$, where $p$ is the probability of each collision type. In the first model, SF (sticking + fragmentation), we study the effect on the fragmentation barrier, in the second, SBF (sticking + bouncing + fragmentation), we study the bouncing barrier, and in the third, SBF+MT (sticking + bouncing + fragmentation + mass transfer), we show a scenario where growth breakthrough occurs.

For the fragmentation, we assume that destruction of both particles always occurs above a given collision velocity
\begin{equation}
p_{\rm f} = \left\{ \begin{array}{ll}
 0 &\mbox{    if $v < v_{\rm f}$} \\
 1 &\mbox{    if $v > v_{\rm f}$} \\
       \end{array} \right.~~~,
\end{equation}
where we take $v_{\rm f} = 100$ cm/s as the fragmentation threshold velocity as found by \cite{1993Icar..106..151B}. The fragments are put in a size-distribution described by $n(m)dm \propto m^{-1.83} dm$.

With bouncing included (where both particles involved are kept unchanged), the sticking efficiency is written as
\begin{equation}
p_{\rm s} = \left\{ \begin{array}{ll}
 1 &\mbox{    if $v < v_{\rm b}$} \\
 0 &\mbox{    if $v > v_{\rm b}$} \\
       \end{array} \right.~~~,
\end{equation}
where we take a bouncing threshold velocity $v_{\rm b} = 5$ cm/s, which is the upper limit to the bouncing threshold velocity found by \cite{2012Icar..218..688W}.

In the collision model SF, we only include sticking and fragmentation, and take $p_{\rm s} = 1 - p_{\rm f}$. In the collision model SBF, we account for all three effects, and write the bouncing probability as $p_{\rm b} = 1 - p_{\rm s} - p_{\rm f}$.

Finally, in the collision model SBF+MT, we include a simple prescription for the mass transfer events discussed in detail in \cite{2012A&A...540A..73W}. These occur when the particle mass ratio is so high that the largest particle can avoid destruction, and only the smaller particle is fragmented with a fraction of its mass being added to the surface of the larger. To mimic this effect, we assume that all fragmenting collisions above a critical mass ratio $m_1/m_2 > 50$ result instead in sticking where 10\% of the mass of the smaller particle is deposited onto the larger one.

\subsection{The velocity distribution}

We assume that all collisions follow a Maxwellian distribution characterized by the root-mean-square velocity $v_{\rm rms}$
\begin{equation}
	\label{eq:maxwellian}
	P(v~|~v_{\rm rms}) = \sqrt{\frac{54}{\pi}} \frac{v^2}{v_{\rm rms}^3} \exp{ \left( -\frac{3}{2} \frac{ v^2}{v_{\rm rms}^2} \right)}~~~.
\end{equation}
For the turbulent relative velocity, we use the closed-form expressions derived by \cite{2007A&A...466..413O}, which for small particles (${\rm St} < 1$, here corresponding to $m \lesssim 10^6$ g) can be approximately written as $v_{\rm rms} = \sqrt{ 9/2 \cdot \alpha~{\rm St}}~c_{\rm s}$ , where St is the Stokes number and the rest of the parameters are given in Table.~\ref{tab:params}.

\begin{figure}[t]
\centering
\resizebox{1.0\hsize}{!}{\includegraphics{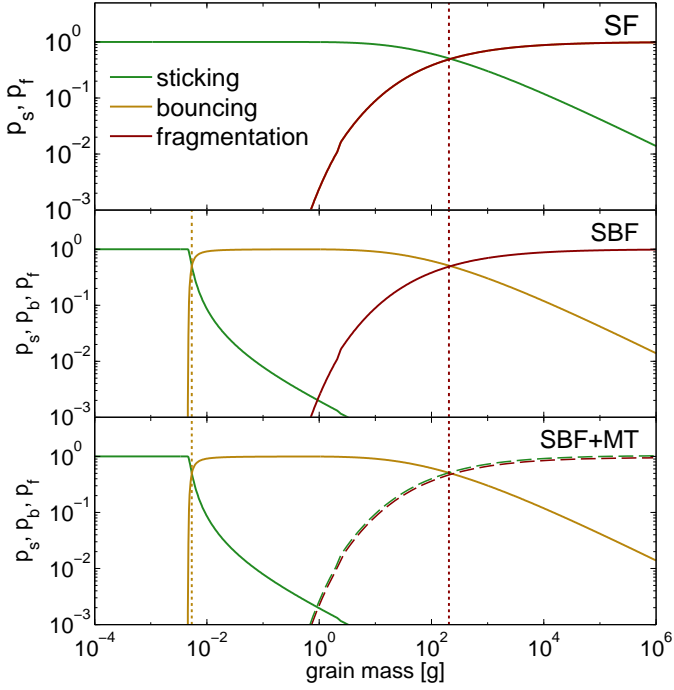}}
\caption{The probabilities of sticking (green), bouncing (yellow), and fragmentation (red) for the three collision models as a function of mass. The red-green dashed line in the SBF+MT panel represents mass transfer, where fragmentation can turn into mass gain for large mass ratios. The dotted lines represent the threshold masses without a velocity distribution included.}
\label{fig:sbfprobs}
\end{figure}

Because of the shape of the velocity distribution, there is always a non-zero, if small, chance of sticking during every collision event. With the collision prescription discussed in the previous section, the total sticking and fragmentation probabilities can be written as
\begin{align}
	\label{eq:stickprob}
	\langle p_{\rm s} \rangle (v_{\rm rms}) &= \int_0^{v_{\rm b}} \! {P(v)}~\mathrm{d}v \\
	\langle p_{\rm f} \rangle (v_{\rm rms}) &= \int_{v_{\rm f}}^{\infty} \! {P(v)}~\mathrm{d}v~~~.
\end{align}
In Fig.~\ref{fig:sbfprobs}, we plot for the three collision models the integrated sticking, bouncing, and fragmentation probabilities as a function of particle mass. In the SF model, fragmentation occurs much earlier if a velocity distribution is included, but sticking is also a possibility at much higher masses. In the SBF model, sticking is a possible outcome even orders of magnitude above the bouncing threshold, but decreases to very low probabilities. At a mass of $m = 1$ g, the sticking probability is $10^{-3}$, but fragmentation is also rare, so that the relative ratio of the two is of the same order of magnitude. This means that growth can still proceed, albeit on longer timescales. Finally, in the SBF+MT model, the situation is identical to the SBF panel, except that the added effect of mass transfer means that collisions above the fragmentation threshold can still lead to growth, provided that the mass ratio between the particles is large enough.

\section{Results}
\label{sec:results}

\begin{figure}[t]
\centering
\resizebox{1.0\hsize}{!}{\includegraphics{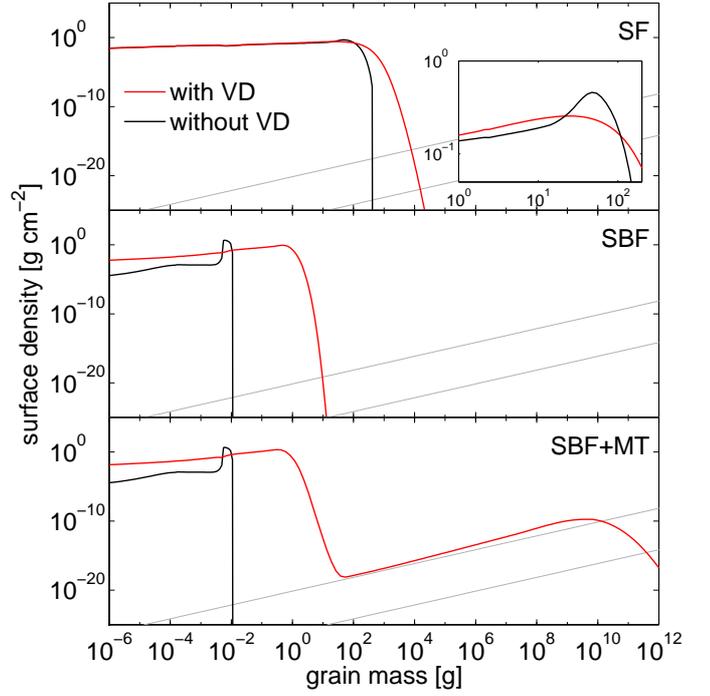}}
\caption{Snapshot of the size-distributions for SF (upper), SBF (middle), and SBF+MT (lower) collision models taken after $t = 5\cdot10^4$ years, with (red) and without (black) a velocity distribution. The gray diagonal lines correspond to a total of 1 and $10^6$ particles within a 0.1 AU annulus, respectively.}
\label{fig:sizedist}
\end{figure}

In Fig.~\ref{fig:sizedist}, the dust size distributions of the simulations are given at $t = 5\cdot10^4$ years for the three collision models, with and without a velocity distribution.

\subsection{The fragmentation barrier}

In the SF model, dust growth stops in both cases. When there is no velocity distribution, this point occurs abruptly, with no way of growing larger particles, after they have reached a size corresponding to $v_{\rm rms} = v_{\rm f}$.

If a probability distribution is considered, growth is both positively and negatively affected. The main peak of the distribution shifts to lower sizes, because collisions from the high-speed tail of the distribution already start their destructive work before the particles reach the barrier (see Fig.~\ref{fig:sizedist} inset). However, there are also lucky particles that successively experience low-velocity collisions, even beyond the nominal barrier. This leads to the tail in the size distribution beyond the barrier seen in Fig.~\ref{fig:sizedist}.

\begin{figure}[t]
\centering
\resizebox{0.9\hsize}{!}{\includegraphics{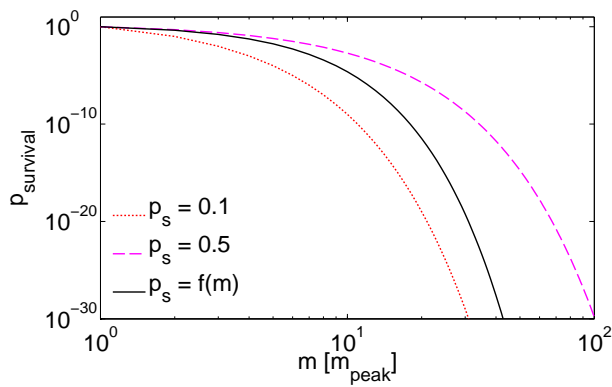}}
\caption{The cumulative survival probability of a particle crossing through the fragmentation barrier. The dotted and dashed lines are calculated by constant $p_{\rm s}$ regardless of particle mass, the solid line by a mass-dependent $p_{\rm s}$ from Eq.~\ref{eq:stickprob}.}
\label{fig:psurvival}
\end{figure}

The probability of a particle reaching a mass $m$ before being destroyed can be approximated by assuming that the particles with masses around the peak mass, $m_{\rm peak}$, dominate the interactions with the larger particles. A particle therefore has to undergo $k$ consecutive sticking collisions in order to grow to a mass of $m = k \cdot m_{\rm peak}$. The survival chance for particles growing from $m_{\rm peak}$ to $k \cdot m_{\rm peak}$ can be written as a product of the sequence
\begin{equation}
	p_{\rm survival} = \prod_{i = 1}^k p_s(m_{\rm i})~~.
\end{equation}
In Fig.~\ref{fig:psurvival}, we plot the cumulative survival probability under different assumptions. The dashed and solid lines were calculated assuming constant sticking probabilities, and the solid line assumed a velocity/mass-dependent $p_{\rm s}$. At a mass of $m = 50 m_{\rm peak}$, the relative velocity had increased by a factor of 2, and the sticking probability had decreased from 0.5 to 0.1 compared to the peak population.

These numbers compare well to the large-particle tail in Fig.~\ref{fig:sizedist}. The largest particles of masses $m = 60 m_{\rm peak}$ have a density decrease of roughly 25 orders of magnitude relative to the peak mass, which is roughly the survival probability that we calculated in the toy-model. Growing to these masses is extremely unlikely, but the sheer number of particles ensures that some lucky particles make it.

\subsection{The bouncing barrier}

In the SBF model, we note two differences between the two cases. One is the discrepancy in the number of small particles. Without a velocity distribution, the small particles are depleted as there is no mechanism that can create them once the main population has grown to the bouncing barrier. This depletion is rapid at first, but gets less effective when the number of small particles drops and the frequency of sticking collisions decreases. If a velocity distribution is included, collisions in the high-velocity tail cause fragmentation and replenish the population of small particles.

The peak of the distribution shifts significantly towards larger sizes. The reason for this can be seen in Fig.~\ref{fig:sbfprobs}, where the low-velocity collisions lead to sticking, but even the highest collision velocities are too low to cause any fragmentation. At around $m = 1$ g, the growth halts when the fragmentation probability increases rapidly, while the sticking collisions are rare. This causes a steeper tail of large particles compared to the case of the pure fragmentation barrier.

We find that bouncing collisions never completely halt the dust growth, as there will always be a small chance of sticking. However, the growth timescale may become so large that the growth is effectively halted. If we follow an individual dust grain, we can write its growth timescale as
\begin{equation}
	\tau_{\rm growth} = \frac{m} {{\rm d} m / {\rm d} t} = \frac{m}{ \sigma~v_{\rm rms}~\rho_{\rm p}~p_{\rm s} }~~~,
\end{equation}
where $\sigma = \pi (a + a_{\rm p})^2$ is the collisional cross-section and $\rho_{\rm p} = 1.4\cdot 10^{-11}$ g cm$^{-3}$ is the midplane mass density of particles that it can collide with. The growth will therefore slow down by a factor proportional to the decrease in $p_{\rm s}$ relative to unhindered coagulation. In the bouncing barrier, this will cause an increase in the growth timescale by a factor of $10^{3}$. Taking $p_{\rm s}$ from Eq.~\ref{eq:stickprob} and the relative velocity prescription of the previous section, we find that it takes $\sim 10,000$ years for particles to grow to $m = 1$ g. If $p_{\rm s}$ were to decrease further, for example owing to a lower bouncing-velocity threshold, this timescale would increase correspondingly.

\subsection{Breaking through the barriers}

In the SBF+MT model, we finally implemented the physics that makes it possible for growth also at high velocities. This relies on a mass difference between the particles in the disk, but without a velocity distribution, such a mass difference never occurs.

With the velocity distribution included, the bouncing barrier can be overcome (see model SBF), and the fragmentation barrier is smoothed out (see model SF), which means that a very small ($10^{6}$ particles in an 0.1 AU annulus) but important fraction of particles manage to grow large enough purely by chance. This triggers the growth of these few lucky particles by sweeping up the smaller grains trapped below the fragmentation barrier.

Even a single fragmenting collision between two lucky particles will create a myriad of fragments that will also be able to sweep up the particles trapped below the barrier. This means that the rare fragmenting collisions will effectively multiply the number of large particles, and with time, even a handful of lucky particles can by themselves create a significant population of planetesimals.

For sweep-up to occur in the simulations, a very high dynamical range is required. Though the break-through occurs for such a tiny fraction of the population, the sweep-up growth causes a rapid increase in both the number and total mass of the larger grains.

\section{Discussion and conclusions}
\label{sec:discussion}

We have found that the collisional growth barriers for dust grains are smoothed out and can even be overcome by virtue of a probability distribution of relative velocities among dust grains. Although improbable, sticky, low-velocity collisions can occur at sizes where the mean collisional velocity would lead to only bouncing or fragmentation.

To grow through the fragmentation barrier, a particle needs to be lucky and experience low-velocity collisions many times in a row, which causes a tail of larger particles to extend from the peak of the mass distribution. Assuming a Maxwellian velocity distribution, the luckiest particles can grow to around 50 times more massive than they would otherwise be.

The bouncing barrier is even more affected by the existence of a velocity distribution, and particles can grow to more than three orders of magnitude higher in mass, with the peak being shifted by two orders of magnitude. This occurs because low-velocity collisions lead to sticking, but even the higher velocities are low enough to only cause bouncing. This means that the growth can continue unimpededly until the average relative velocities have increased enough for the fragmentation barrier to start to become important. The bouncing barrier is therefore not a solid barrier at all, unless the growth timescale becomes too long because of the low sticking probability.

We have found that the low-velocity tail allows some lucky particles to grow beyond the bouncing and fragmentation barriers, to become the first seeds in the sweep-up scenario introduced by \cite{2012A&A...540A..73W}. When the effect of fragmentation-with-mass-transfer is included, these seeds can sweep up the smaller particles trapped by the growth barriers, and start their growth towards planetesimal sizes.

\begin{acknowledgements}
We would like to thank the referee for helpful comments. F.W. was funded by the Deutsche Forschungsgemeinschaft within the Forschergruppe 759 "The Formation of Planets: The Critical First Growth Phase". For C.W.O. support for this
work was provided by NASA through Hubble Fellowship grant \#HST-HF-51294.01-A awarded by the Space Telescope Science Institute, operated by the Association of Universities for Research in Astronomy, Inc., for NASA, under contract NAS 5-26555.
\end{acknowledgements}

\bibliographystyle{aa}
\bibliography{refs}

\makeatletter
\if@referee
\processdelayedfloats
\pagestyle{plain}
\fi
\makeatother
\end{document}